\documentclass[letterpaper,twocolumn,english,conference]{IEEEtran}
\usepackage{mathptmx}

\usepackage[T1]{fontenc}
\usepackage[latin9]{inputenc}
\usepackage{color}
\usepackage{units}
\usepackage{mathtools}
\usepackage{algorithm2e}
\usepackage{amsmath}
\usepackage{amsthm}
\usepackage{amssymb}
\usepackage{graphicx}
\PassOptionsToPackage{normalem}{ulem}
\usepackage{ulem}

\makeatletter

\pdfpageheight\paperheight
\pdfpagewidth\paperwidth

\providecommand{\tabularnewline}{\\}

\theoremstyle{plain}
\newtheorem{thm}{\protect\theoremname}
\theoremstyle{remark}
\newtheorem{rem}[thm]{\protect\remarkname}

\usepackage[hang,flushmargin]{footmisc}
\usepackage{babel}
\usepackage[font={small}]{caption}
\usepackage{cite}
\usepackage{flushend}
\LinesNumbered
\SetKwRepeat{Do}{do}{while}%

\renewcommand{\fnum@figure}{Fig. \thefigure}

\providecommand{\remarkname}{Remark}
\providecommand{\theoremname}{Theorem}

\usepackage{microtype}

\providecommand{\remarkname}{Remark}
\providecommand{\theoremname}{Theorem}

\usepackage{babel}
\providecommand{\remarkname}{Remark}
\providecommand{\theoremname}{Theorem}

\makeatother

\usepackage{babel}
\providecommand{\remarkname}{Remark}
\providecommand{\theoremname}{Theorem}

\begin{document}

\title{Realizing Wireless Communication through Software-defined HyperSurface
Environments}

\author{Christos Liaskos\IEEEauthorrefmark{1}, Shuai Nie\IEEEauthorrefmark{2},
Ageliki Tsioliaridou\IEEEauthorrefmark{1}, Andreas Pitsillides\IEEEauthorrefmark{3},
Sotiris Ioannidis\IEEEauthorrefmark{1}, and Ian Akyildiz\IEEEauthorrefmark{2}\IEEEauthorrefmark{3}\\
 {\small{}\IEEEauthorrefmark{1}Foundation for Research and Technology
- Hellas (FORTH)}\\
 {\small{}Emails: \{cliaskos,atsiolia,sotiris\}@ics.forth.gr}\\
 {\small{}\IEEEauthorrefmark{2}Georgia Institute of Technology, School
of Electrical and Computer Engineering}\\
 {\small{}Emails: \{shuainie, ian\}@ece.gatech.edu}\\
 {\small{}\IEEEauthorrefmark{3}University of Cyprus, Computer Science
Department}\\
 {\small{}Email: Andreas.Pitsillides@ucy.ac.cy}\\
}
\maketitle
\begin{abstract}
Wireless communication environments are unaware of the ongoing data
exchange efforts within them. Moreover, their effect on the communication
quality is intractable in all but the simplest cases. The present
work proposes a new paradigm, where indoor scattering becomes software-defined
and, subsequently, optimizable across wide frequency ranges. Moreover,
the controlled scattering can surpass natural behavior, exemplary
overriding Snell's law, reflecting waves towards any custom angle
(including negative ones). Thus, path loss and multi-path fading effects
can be controlled and mitigated. The core technology of this new paradigm
are metasurfaces, planar artificial structures whose effect on impinging
electromagnetic waves is fully defined by their macro-structure. The
present study contributes the software-programmable wireless environment
model, consisting of several HyperSurface tiles controlled by a central,
environment configuration server. HyperSurfaces are a novel class
of metasurfaces whose structure and, hence, electromagnetic behavior
can be altered and controlled via a software interface. Multiple networked
tiles coat indoor objects, allowing fine-grained, customizable reflection,
absorption or polarization overall. A central server calculates and
deploys the optimal electromagnetic interaction per tile, to the benefit
of communicating devices. Realistic simulations using full 3D ray-tracing
demonstrate the groundbreaking potential of the proposed approach
in~$2.4\,GHz$ and $60\,GHz$ frequencies.
\end{abstract}

\begin{IEEEkeywords}
Wireless Environment, Communication-awareness, Indoor, Millimeter
wave, Software control, Metasurfaces.
\end{IEEEkeywords}

\section{Introduction\label{sec:Intro}}

Recent years have witnessed a tremendous increase in the efficiency
of wireless communications. Multiple techniques have been developed
to tackle the stochastic nature of the wireless channel, in an effort
to fully adapt to its wide fluctuations. Indoor environments have
attracted special attention, since multi-path fading accentuates due
to the presence of multiple scatterers in a confined space. In such
cases, techniques such as MIMO, beamforming, adaptive modulation and
encoding have enabled wireless devices to rapidly adapt to the time-variant,
unpredictable channel state~\cite{pi2016millimeter}. The present
work opens an unexplored research path: making the wireless environment
fully controllable via software, enabling the optimization of major
propagation factors between wireless devices. Thus, effects such as
path loss and multi-path fading become controllable and mitigate-able.

In order to understand the potential of exerting control over an environment,
we first need to define its composition and its natural behavior.
Indoor environments, which constitute the focus of the present work,
comprise two or more communicating devices\textendash such as laptops,
mobile phones, access points, base stations etc.\textendash and any
object found in a domestic or work space that can influence their
communication. At lower frequencies, walls, ceilings, floors, doors
and sizable furniture act as electromagnetic (EM) wave scatterers,
creating multiple paths between communicating end-points, especially
in non-line-of-sight (NLOS) areas. At higher frequencies, such as
millimeter wave (mm-wave) or terahertz ($THz$), which are expected
to play a major role in upcoming 5G communications~\cite{yilmaz2016millimetre},
even small objects act as substantial scatterers. Furthermore, ultra-small
wavelengths translate to considerable Doppler shift even at pedestrian
speed~\cite{yilmaz2016millimetre}. These factors, coupled with the
natural ambient dissipation of power due to free space losses, lead
to undermined NLOS performance at $2-5\,GHz$ and inability for NLOS
communications at $60\,GHz$ and beyond~\cite{yilmaz2016millimetre}.

Existing proposals for multi-path and path loss mitigation can be
classified as i) device-oriented, and ii) retransmitter-oriented.
Device-oriented methods include massive MIMO deployments in communicating
devices, to make constructive use of the multi-path phenomena~\cite{Aijaz.2017}.
Additionally, beamforming seeks to adaptively align the direction
of wireless transmissions in order to avoid redundant free space losses~\cite{kelif20163d,REFLECTARRAYS}.
Additional schemes include the on-the-fly selection of the modulation
and encoding scheme that offers the best bit error rate (BER) under
the current channel conditions~\cite{huang2017multi}. Retransmission-oriented
solutions advocate for the placement of amplifiers in key-positions
within the indoor environment. Retransmitters can be either passive
or active: Passive retransmitters are essentially conductive structures
akin to antenna plates~\cite{reflectInfocom.2017}. They passively
reflect energy from and towards fixed directions, without tunability.
Active retransmitters are powered electronic devices that amplify
and re-transmit received signals within a given frequency band. Essentially,
they attempt to combat power loss by diffusing more power within the
environment. In mm-wave frequencies and beyond, retransmitters must
be placed in line-of-sight (LOS) among each other, in an effort to
eliminate NLOS areas within a floor plan. Device-to-device networking
can also act as a retransmission solution for specific protocols and
a limited capacity of served users~\cite{chen2017promoting}. The
overviewed solutions have a common trait: They constitute device-side
approaches, which treat the environment as an uncontrollable factor
that does not participate into the communication process.

Metasurfaces are the core technology for introducing programmatically
controlled wireless environments~\cite{Zhu.2017,Minovich.2015,Lucyszyn.2010}.
They constitute the outcome of a research direction in Physics interested
in creating (rather than searching for) materials with required EM
properties. In their earlier iterations, they comprised a metallic
pattern, called \emph{meta-atom}, periodically repeated over a Silicon
substrate, as shown in Fig.~\ref{fig:Mspatterns}. The macroscopic
EM behavior of a metasurface is fully defined by the meta-atom form.
A certain pattern may fully absorb all impinging EM waves from a given
direction of arrival (DoA), while another may fully reflect a given
DoA towards another, at a negative reflection angle. Notably, metasurfaces
(and their 3D counterpart, the metamaterials) offer a superset of
EM behaviors with regard to regular materials. Lens functionality
(concentration of reflections towards a given point rather than ambient
dispersal) and negative refraction/reflection indexes are but some
of the exotic EM capabilities they can exhibit~\cite{Chen.2016}.
Dynamic meta-atom designs allow for dynamic metasurfaces, as shown
in Fig.~\ref{fig:Mspatterns}. Such designs include tunable factors,
such as CMOS switches or Micro Electro-Mechanical Switches (MEMS)
that can alter their state\textendash and the EM behavior of the metasurface\textendash via
an external bias~\cite{Chen.2016}. The bias is commonly electronic,
but thermal, light-based and mechanical approaches have been studied
as well~\cite{Chen.2016}. Thus, multi-functional metasurfaces, that
can switch from one EM behavior to another (e.g., from absorbing to
custom steering) are enabled. Finally, a very strong trait is that
there is no known limitation to the operating metasurface frequency,
which can be at the $mm$-wave and $THz$ bands~\cite{Lee.2012}.

\begin{figure}[t]
\begin{centering}
\includegraphics[width=1\columnwidth]{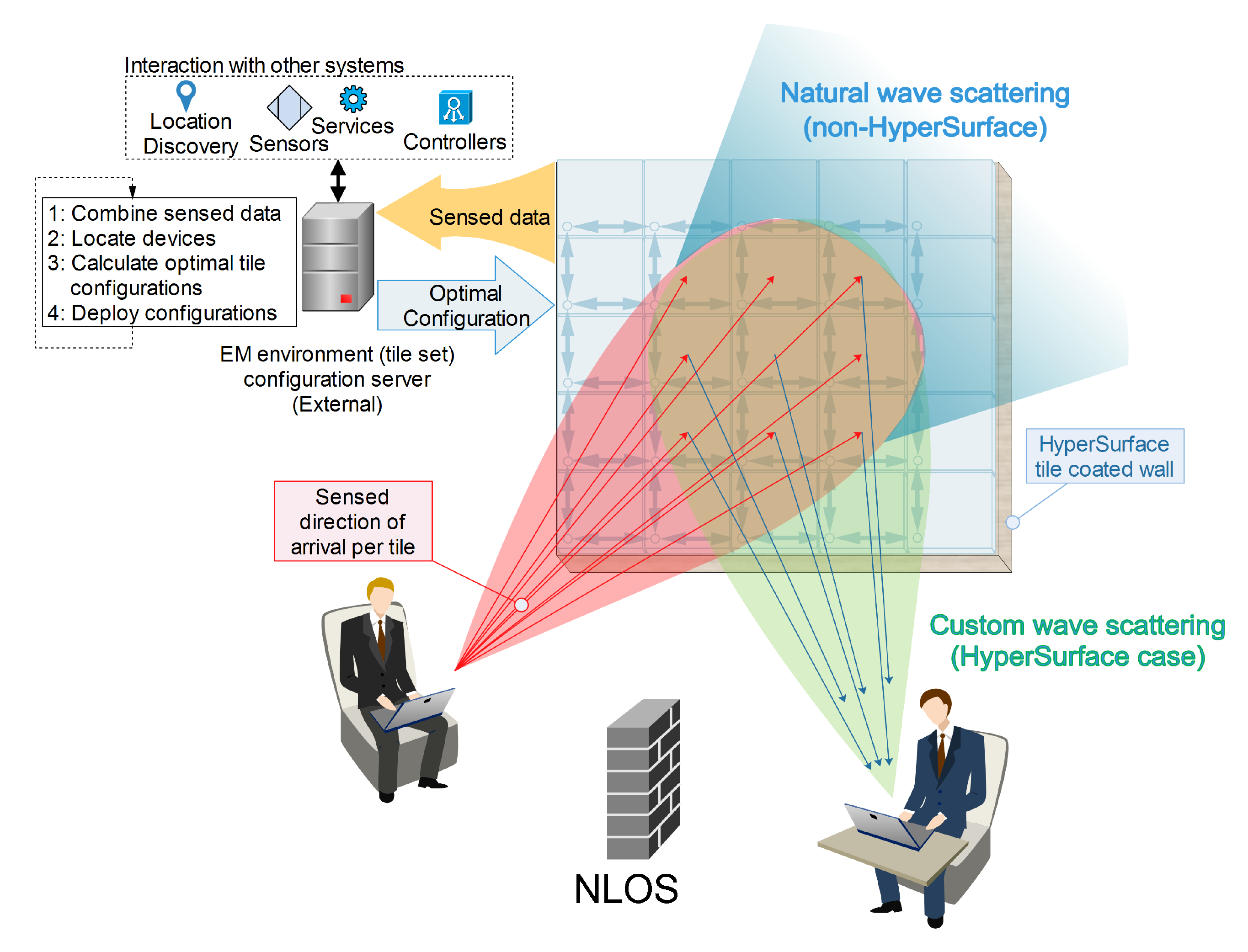}
\par\end{centering}
\caption{\label{fig:workflow}The proposed workflow involving HyperSurface
tile-coated environmental objects. The EM scattering is tailored to
the needs of the communication link under optimization. Unnatural
EM scattering, such as lens-like EM focus and negative reflection
angles can be employed to mitigate path loss and multi-path phenomena,
especially in challenging NLOS cases.}
\end{figure}

The methodology proposed by the present study is to coat objects of
EM significance within an indoor environment with a novel class of
software-controlled metasurfaces. The study defines a unit of this
metasurface class, called \emph{HyperSurface tile}. A HyperSurface
tile is a planar, rectangular structure that incorporates networked
hardware control elements and adaptive meta-atom metasurfaces. Following
a well-defined programming interface, a tile can receive external
commands and set the states of its control elements to match the intended
EM behavior. The tiles, covering walls, doors, offices, etc., form
networks to facilitate the relaying of programmatic commands among
them. Moreover, tiles can have environmental sensing and reporting
capabilities, facilitating the discovery of communicating devices
within the environment. As shown in Fig.~\ref{fig:workflow}, a central
server can receive incoming tile reports, calculate the optimal configuration
per tile, and set the environment in the intended state by sending
the corresponding commands. Collaboration with existing systems (e.g.,
localization services and Cloud computing), constitutes a strong aspect
of the proposed approach, given that it enables the incorporation
of the EM behavior of materials in smart control loops.

The present study contributes the first model to describe programmable
wireless indoor environments, detailing their hardware, networking
and software components. The model includes the way for translating
EM metasurface functionalities to reusable software functions, bridging
physics and informatics. Moreover, the protocol specifications and
programming interfaces for interacting with tiles for communication
purposes are outlined. The practical procedure for deploying programmable
EM environments to mm-wave indoor communication is detailed. The potential
of programmable environments is evaluated via full 3D ray tracing
in $2.4$ and $60\,GHz$ cases, demonstrating their ground-breaking
potential in mitigating path loss and multi-path fading effects. The
study also presents a way of modifying common ray-tracers to enable
their use in programmable EM environment simulations. Uses of the
proposed concept in other application domains are also discussed.

The remainder of the text is organized as follows. Prerequisite knowledge
on metasurfaces is given in Section~\ref{sec:Backgrd}. The HyperSurface-based
wireless environment model is given in Section~\ref{sec:arch}. Applications
to indoor wireless setups are discussed in Section~\ref{sec:App}.
Evaluation via ray-tracing-based simulations is presented in Section~\ref{sec:Evaluation}.
Finally, the conclusion is given in Section~\ref{sec:Conclusion}.

\section{Prerequisites \label{sec:Backgrd}}

This section provides the necessary background knowledge on metasurfaces,
discussing dimensions and composition, operating principles and supported
functionalities. The following concise description targets a wireless
communications audience, given the topic of the present paper. A more
detailed introduction can be found in~\cite{Banerjee.2011}.

A metasurface is a planar, artificial structure which comprises a
repeated element, the meta-atom, over a substrate. In most usual compositions,
the meta-atom is conductive and the substrate is dielectric. Common
choices are copper over silicon, while silver and gold constitute
other exemplary conductors~\cite{Zhu.2017}. More exotic approaches
employ graphene, in order to interact with $THz$-modulated waves~\cite{Lee.2012}.
Metasurfaces are able to control EM waves impinging on them, in a
frequency span that depends on the overall dimensions. The size of
the meta-atom is comparable to the intended interaction wavelength,~$\lambda$,
with $\nicefrac{\lambda}{2}$ constituting a common choice. The thickness
of the metasurface is smaller than the interaction wavelength, ranging
between $\nicefrac{\lambda}{10}\to\nicefrac{\lambda}{5}$ as a rule
of a thumb. Metasurfaces usually comprise several hundreds of meta-atoms,
which results into fine-grained control over the EM interaction control.
In general, a minimum size of approximately $30\times30$ meta-atoms
is required to yield an intended EM interaction~\cite{Chen.2016}.
\begin{figure}[t]
\begin{centering}
\includegraphics[width=1\columnwidth]{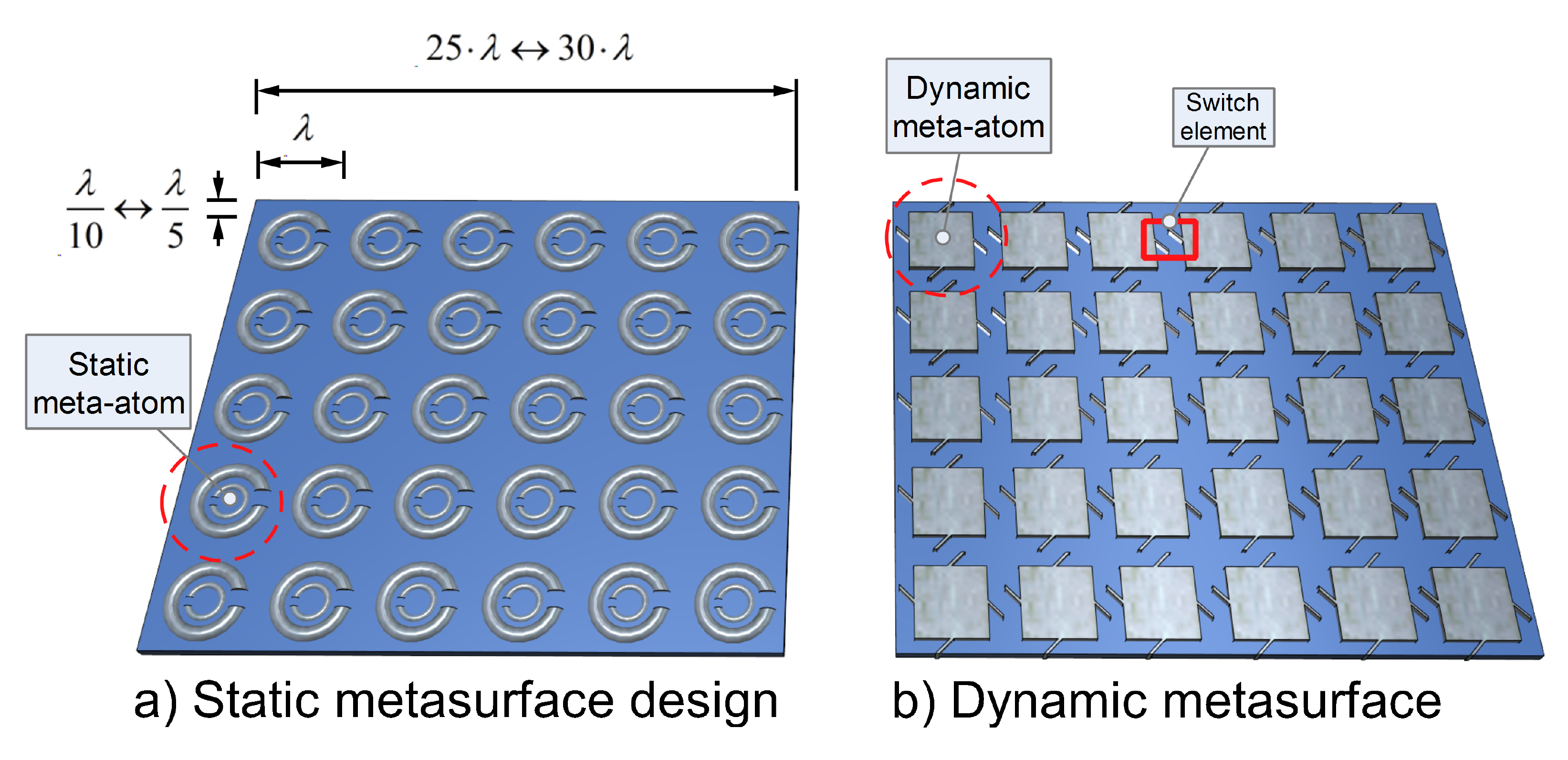}
\par\end{centering}
\caption{\label{fig:Mspatterns}Split ring resonators (left) constituted a
very common type of static metasurfaces, with fixed EM behavior. Novel
designs (right) incorporate switch elements (MEMS, CMOS or other)
to offer dynamically tunable EM behavior.}
\end{figure}

Figure~\ref{fig:Mspatterns}-a illustrates a well-studied metasurface
design comprising split-ring resonators as the meta-atom pattern.
Such classic designs that rely on a static meta-atom, naturally yield
a static interaction with EM waves. The need for dynamic alteration
of the EM wave control type has given rise to dynamic metasurfaces,
illustrated in Fig.~\ref{fig:Mspatterns}-b. Dynamic meta-atoms incorporate
phase switching components, such as MEMS or CMOS transistors, which
can alter the structure of the meta-atom. Thus, dynamic meta-atoms
allow for time-variant EM interaction, while meta-atom alterations
may give rise to multi-frequency operation~\cite{Zhu.2017}. Phase
switching components can also be classified into state-preserving
or not. For instance, mechanical switches may retain their state and
require powering only for state transitions, while semiconductor switches
require power to maintain their state.

\begin{figure}[t]
\begin{centering}
\includegraphics[width=1\columnwidth]{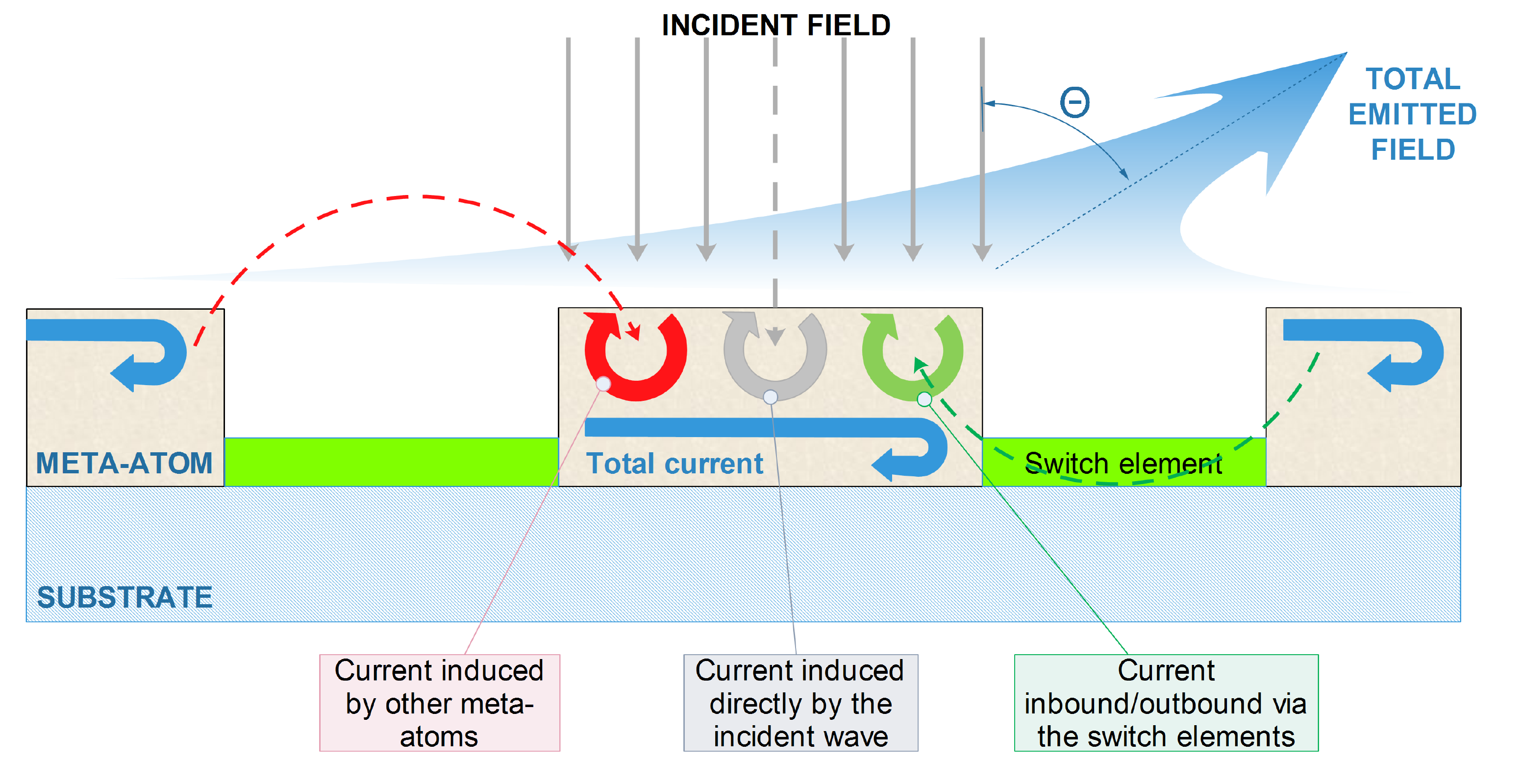}
\par\end{centering}
\caption{\label{fig:MSprinciple}The principle of metasurface functionality.
Incident waves create a well-defined EM response to the unit cells.
The cell response is crafted in such a way that the aggregate field
follows a metasurface-wide design objective, e.g., reflection towards
a custom angle $\Theta$.}
\end{figure}

The operating principle of metasurfaces is given in Fig.~\ref{fig:MSprinciple}.
The meta-atoms, and their interconnected switch elements in the dynamic
case, act as control factors over the surface currents flowing over
the metasurface. The total EM response of the metasurface is then
derived as the total emitted field by all surface currents, and can
take completely engineered forms, such as the unnatural reflection
angle shown in Fig.~\ref{fig:MSprinciple}. Engineering the total
surface current is a complex process that must account for currents
directly induced over the metasurface by the incident wave, the currents
induced in a meta-atom wirelessly by other meta-atoms, as well as
the currents flowing inwards or outwards from a meta-atom via the
switch elements. A qualitative description of the dynamic metasurface
operation can also be given: the meta-atoms can be viewed as either
input or output antennas, connected in custom topologies via the switch
elements. Impinging waves enter from the input antennas, get routed
according to the switch element states, and exit via the output antennas,
exemplary achieving customized reflection.

\subsection{State-of-the-art potential and manufacturing approaches}

Metasurfaces constitute the state of the art in EM control in terms
of capabilities and control granularity. A metasurface can support
a wide range of EM interactions, denoted as \emph{functions}. Common
function types include~\cite{Minovich.2015}:
\begin{itemize}
\item Reflection of an impinging wave, with a given direction of arrival,
towards a completely custom direction.
\item Refraction of EM waves via the metasurface towards any inwards direction.
Both the reflection and refraction functions can override the outgoing
directions predicted by Snell's law. Reflection and refraction functions
will jointly be referred to as wave \emph{steering}.
\item Wave absorbing, i.e., ensuring minimal reflected and/or refracted
power for impinging waves.
\item Wave polarizing, i.e., changing the oscillation orientation of the
wave's electric and magnetic field.
\end{itemize}
Moreover, they can offer additional, advanced functions, such as collimation,
resulting from near zero permittivity and permeability response, anisotropic
response leading to hyperbolic dispersion relation, giant chirality,
arbitrary wave-front shaping and frequency selective filtering~\cite{Lucyszyn.2010}.
Apart from communications, these traits have been exploited in a variety
of applications, e.g., highly efficient energy harvesting photovoltaics,
and thermophotovoltaics, ultra-high resolution medical imaging, sensing,
quantum optics and military applications~\cite{Iwaszczuk.2012}.

The extended repertoire of EM function types, as well the exquisite
degree of granularity in EM behavior control, sets metasurfaces apart
from phased antennas and reflectarrays~\cite{PHASEDANTENNAS,REFLECTARRAYS},
which can only support coarse EM steering, e.g., for beamforming applications
in wireless devices~\cite{kelif20163d}. Notice that highly fine-grained
EM control is required in mm-wave setups, due to the extremely small
wavelength~\cite{yilmaz2016millimetre}.

Regarding their manufacturing approaches, metasurfaces are common
produced as conventional printed circuit boards (PCBs)~\cite{Yang.2016}.
The PCB approach has the advantage of relying on a mature, commercially
accessible manufacturing technology. The PCB production cost is moderate
(indicatively, USD~500 per $m^{2}$~\cite{PCBCART}). However, the
PCB technology is originally intended for integrated circuits with
far greater complexity than a metasurface. As described in the context
of Fig.~\ref{fig:Mspatterns}, a metasurface can be a very simple
structure, comprising a set of conductive patches, diodes and conductive
power/signal lines. Therefore, large area electronics (LAE) can constitute
better manufacturing approaches in terms of ultra low production cost~\cite{LAEBOOK,LAEAPP}.
LAE can be manufactured using conductive ink-based printing methods
on flexible and transparent polymer films, and incorporate polymer/organic
diodes~\cite{LAEPRINTED}. Films with metasurface patterns and diodes
printed on them can then be placed upon common objects (e.g., glass,
doors, walls, desks), which may also act as the dielectric substrate
for the metasurface.

\section{The HyperSurface-based Wireless Environment Model\label{sec:arch}}

\begin{figure*}[t]
\begin{centering}
\includegraphics[width=1\textwidth]{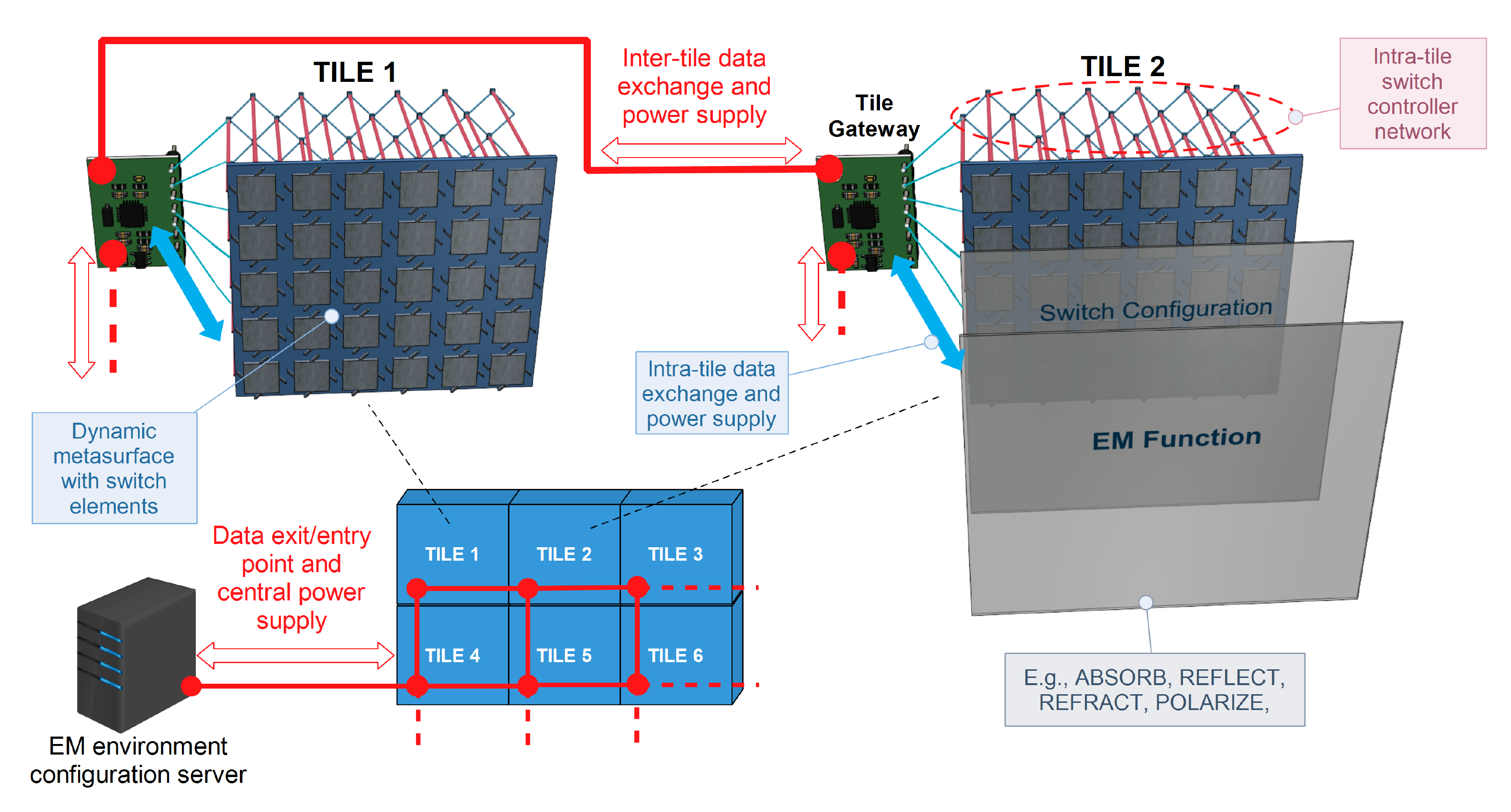}
\par\end{centering}
\caption{\label{fig:Architecture}Illustration of the HyperSurface tile architecture
and the tile-enabled wireless environment model. }
\end{figure*}

This section details the HyperSurface tile hardware components, the
tile inter-networking and the environment control software. A schematic
overview is given in Fig.~\ref{fig:Architecture} and is detailed
below.

\uline{The tile hardware}. The tile hardware consists of a dynamic
metasurface, a set of networked, miniaturized controllers that control
the switch elements of the metasurface, and a gateway that provides
inter-tile and external connectivity. The controller network has a
slave/master relation to the gateway. Via the gateway, the controller
network reports its current state and receives commands to alter the
state of the switch elements in a robust manner, making the metasurface
yield an overall required EM function.

A single controller is a miniaturized, addressable electronic device
that can monitor and modify the state of at least one metasurface
switch element. The controller design objectives are small size (to
avoid significant interference to the EM function of the metasurface),
low-cost (to support massive deployments in many tiles), high monitoring
and actuation speed (to sustain fast EM reconfigurability of the metasurface),
and the ability to create, receive and relay data packets (to enable
controller networking).

The avoidance of EM function disruption also refers to the wiring
required to connect the controllers to the switch elements and to
each other in a grid topology (cf. Fig.~\ref{fig:Architecture}).
Therefore, the total wiring should also be kept low. The grid-networked
controller approach is an option that balances wiring length and robustness
to node failures. Bus connectivity for the controllers would minimize
the required wiring, but would decrease the robustness against node
failures. On the other hand, a star connectivity would offer maximum
robustness but would also yield maximum wiring. Notice that future
technologies, such as nanonetworking, may enable wireless, computationally-powerful
nodes with autonomous, energy harvesting-based power supply~\cite{Akyildiz.2008}.
Thus, future tile designs may need no wiring or specific gateways.
The setup presented in this study prioritizes cost-effective realizability
with present manufacturing capabilities.

At a logical level, a controller is modeled as a finite-state automaton,
which reacts to incoming packets or switch element changes by transitioning
from one state to another~\cite{books2011automata}. A UML-standard
state diagram should capture three basic controller processes: the
data packet handling (including re-routing, consuming packets and
sending acknowledgments), the node reporting (reacting to an incoming
monitoring directive\textendash monitor request packet\textendash by
creating a new monitor data packet), and a fault detection process
(either self- or neighbor-failure). The latter is required for robust
data routing and for deducing the operational state of the tile as
a whole. Regarding the controller addressing, it can be hardwired
due to the fixed grid topology.

The tile gateway stands between the tile controller network and the
external world. It is incorporated to the tile fabric at a position
selected to yield minimal EM interaction concerns (e.g., at the back
of the tile). It provides mainstream protocol-compatible data exchange
with any other system. Internally, it is connected to at least one
controller, while more connections can be used for robust connectivity.
Moreover the gateway acts as a power supply bridge for the tile. Limited
size (e.g.,~\textasciitilde{}$cm$) and energy requirements are the
only significant constraints. Existing hardware, such as IoT platforms~\cite{Verikoukis.2017},
can be employed as tile gateways~\cite{Verikoukis.2017}. The tile
gateway may optionally have EM DoA sensing capabilities, to facilitate
the location discovery of wireless user devices in the environment.

\uline{The tile inter-networking}. As tiles are placed over an
environmental object, such as a wall, they click together, connecting
data and power lines among the tile gateways (cf. Fig.~\ref{fig:Architecture}).
Thus, the tiles form a wired ad hoc network in a grid topology. Once
again, existing IoT communication protocols can be readily employed.
The same protocol is used for connecting the tile network to any external
system. At least one tile\textendash denoted as exit/entry point\textendash has
its gateway connect to the environment configuration server, which
accumulates sensed data and diffuses EM actuation commands within
the tile network. More than one tile can be used as exit/entry points
at the same time, for the interest or robust and timely data delivery.

\uline{The environment control software}. The environment control
software is an application programming interface (API) that exists
at the configuration server. The API serves as a strong layer of abstraction,
hiding the internal complexity of the HyperSurfaces. It offers user-friendly
and general purpose access to metasurface functions, without requiring
knowledge of the underlying hardware and Physics. It provides software
descriptions of metasurface functions, allowing a programmer to customize,
deploy or retract them on-demand over tiles with appropriate callbacks.
These callbacks have the following general form:
\[
{\scriptstyle \texttt{outcome}\gets\texttt{callback(tile\_ID, action\_type, parameters)}}
\]
The $\texttt{tile\_ID}$ is the unique address of the intended tile
gateway in the inter-tile network (e.g., an IPv6). One EM function
per tile is considered here for simplicity. The $\texttt{action\_type}$
is an identifier denoting the intended function, such as $\texttt{STEER}$
or $\texttt{ABSORB}$, as described in Section~\ref{sec:Backgrd}.
Each action type is associated to a set of valid parameters. For instance,
$\texttt{STEER}$ commands require: i) an incident DoA, $\overrightarrow{I}$,
ii) an intended reflection direction, $\overrightarrow{O}$, and iii)
the applicable wavelength, $\lambda$, (if more than one are supported).
$\texttt{ABSORB}$ commands require no $\overrightarrow{O}$ parameter.
Notice that metasurface properties can be symmetric: i.e., a $\texttt{STEER}\left(\overrightarrow{I},\overrightarrow{O}\right)$
can also result into $\texttt{STEER}\left(\overrightarrow{O},\overrightarrow{I}\right)$~\cite{Holloway.2012}.

Once executed at the configuration server, a callback is translated
to an appropriate configuration of the switch elements that should
be deployed at the intended tile. The configuration is formatted as
a data packet that enters the tile network via an entry/exit point,
and is routed to the intended tile via the employed inter-tile routing
protocol. The intended tile gateway translates the directive according
to the controller network communication protocol specifications and
diffuses it within the tile. Upon success, it returns an acknowledgment
to the configuration server, or an error notification otherwise.

In the general case, the translation of an EM function to a tile switch
element configuration is accomplished via a lookup table, populated
during the tile design/manufacturing process as follows. Let $\sigma$
be a single tile configuration, defined as an array with elements
$s_{ij}$ describing the intended switch element state that is overlooked
by controller with address $i,j$ in the tile controller network.
(One-to-one controller-switch relation is assumed). In the MEMS case,
$s_{ij}$ takes binary values, $1$~or~$0$, denoting switch connection
or disconnection. Additionally, let $\Sigma$ be the set of all possible
configurations, i.e., $\sigma\in\Sigma$. Let an EM function of type
$\texttt{ABSORB}$ from DoA $\overrightarrow{I}$ be of interest.
Moreover, let $P_{\sigma}(\phi,\theta)$ be the power reflection pattern
of the tile (in spherical coordinates), when a wave with DoA $\overrightarrow{I}$
impinges upon it and a configuration $\sigma$ is active. Then, the
configuration $\sigma_{best}$ that best matches the intended function
$\texttt{ABSORB}\left(\overrightarrow{I}\right)$ is defined as:
\begin{equation}
\sigma_{best}\gets argmin_{\sigma\in\Sigma}\left\{ max_{\forall\phi,\theta}P_{\sigma}(\phi,\theta)\right\} \label{eq:fitness}
\end{equation}
Existing heuristic optimization processes can solve this optimization
problem for all functions of interest in an offline manner~\cite{haupt2007genetic},
using simulations or field measurements on prototypes. The configuration
lookup table is thus populated. Finally, we note that analytical results
for the EM function-configuration relation exist in the literature
for several metasurface designs~\cite{haupt2007genetic}. In such
cases, the analytical results can be employed directly, without the
need for lookup tables.

\section{Control Algorithm of HyperSurface\label{sec:algorithm}}

In order to establish communication links between transmitters and
receivers, the HyperSurface tiles need to be adaptively selected and
optimally controlled to serve the desired receivers. Since in real-world
communication scenarios, multiple users can be present in the same
space, it is necessary to discuss the tile distribution and control
algorithms.

\textcolor{black}{We start from the case where there is one pair of
transmitter and receiver in the environment, as shown in Fig.~\ref{fig:WirCommExampleNegAngle}.
When the transmitter sends signals, multiple tiles can sense the transmitted
signals. According to the location of the receiver sensed by the location
discovery system, those tiles will steer their angles to establish
reflection paths. Therefore, the signal received at the receiver is
a superposition of all signals reflected from various tiles. Among
all tiles, only those that can sense the transmitted signals will
respond to forwarding requests and tune their angles. On the other
hand, in a more complicated case where multiple users are present
in the same environment, the tile distribution needs to be optimized.
The signals to be received by different users should be orthogonal
to each other and are forwarded by different HyperSurface paths.}

\textcolor{black}{We assume the transmitted signal is QPSK modulated
with symbol $k(t)$, thus in general the received signal in time domain
can be expressed as
\begin{equation}
r(t)=k(t)\sum_{i=1}^{N}a_{i}e^{-j\theta_{i}}e^{j2\pi f_{c}\tau_{i}}+n(t),
\end{equation}
where $f_{c}$ is the central frequency, $a_{i}$, $\theta_{i}$,
and $\tau_{i}$ are the attenuation, phase, and delay caused by the
reflection paths along HyperSurface tiles of $i$-th path, and $n(t)$
is the AWGN noise in the channel. We assume there are in total $N$
paths found between the transmitter and receiver. The multipath effects
might cause distortion in overall received signal, therefore we need
to mitigate the destructive interference and harmonize the phases
by controlling the operation of HyperSurface tiles. Specifically,
we can formulate it as an optimization problem aimed at maximizing
the received power, $P_{r}^{(j)}$, and the number of tiles of the
HyperSurface, $M_{HS}^{(j)}$, for the $j^{\mathrm{th}}$ receiver
in the network with a total of $J$ users with $d_{j}$ distance,
as follows:
\begin{align}
\textit{Given:} & (x_{t},y_{t},z_{t}),(x_{r}^{(j)},y_{r}^{(j)},z_{r}^{(j)}),\\
 & P_{t}^{\mathrm{total}},M_{HS}^{\mathrm{total}}\\
\textit{Find: } & P_{t}^{(j)},M_{HS}^{(j)}\\
\textit{Objective: } & \max\sum d_{j}P_{r}^{(j)}\\
\textit{Subject to: }\nonumber \\
\text{Transmit power allocation: } & \sum P_{t}^{(j)}\leq P_{t}^{\mathrm{total}}\\
\text{HyperSurface tile allocation: } & \sum M_{HS}^{(j)}\leq M_{HS}^{\mathrm{total}}\\
\text{ for all }j\in J
\end{align}
In the above optimization problem, $(x_{t},y_{t},z_{t})$ and $(x_{r}^{(i)},y_{r}^{(i)},z_{r}^{(i)})$
denote the three-dimensional coordinates of the transmitter and the
$i^{\mathrm{th}}$ receiver, respectively. Based on the above optimization
problem, we can distribute tiles to corresponding users without causing
interference or signal distortion. }

\section{Applications to Mm-wave Indoor Setups\label{sec:App}}

\begin{figure}[t]
\begin{centering}
\includegraphics[width=1\columnwidth]{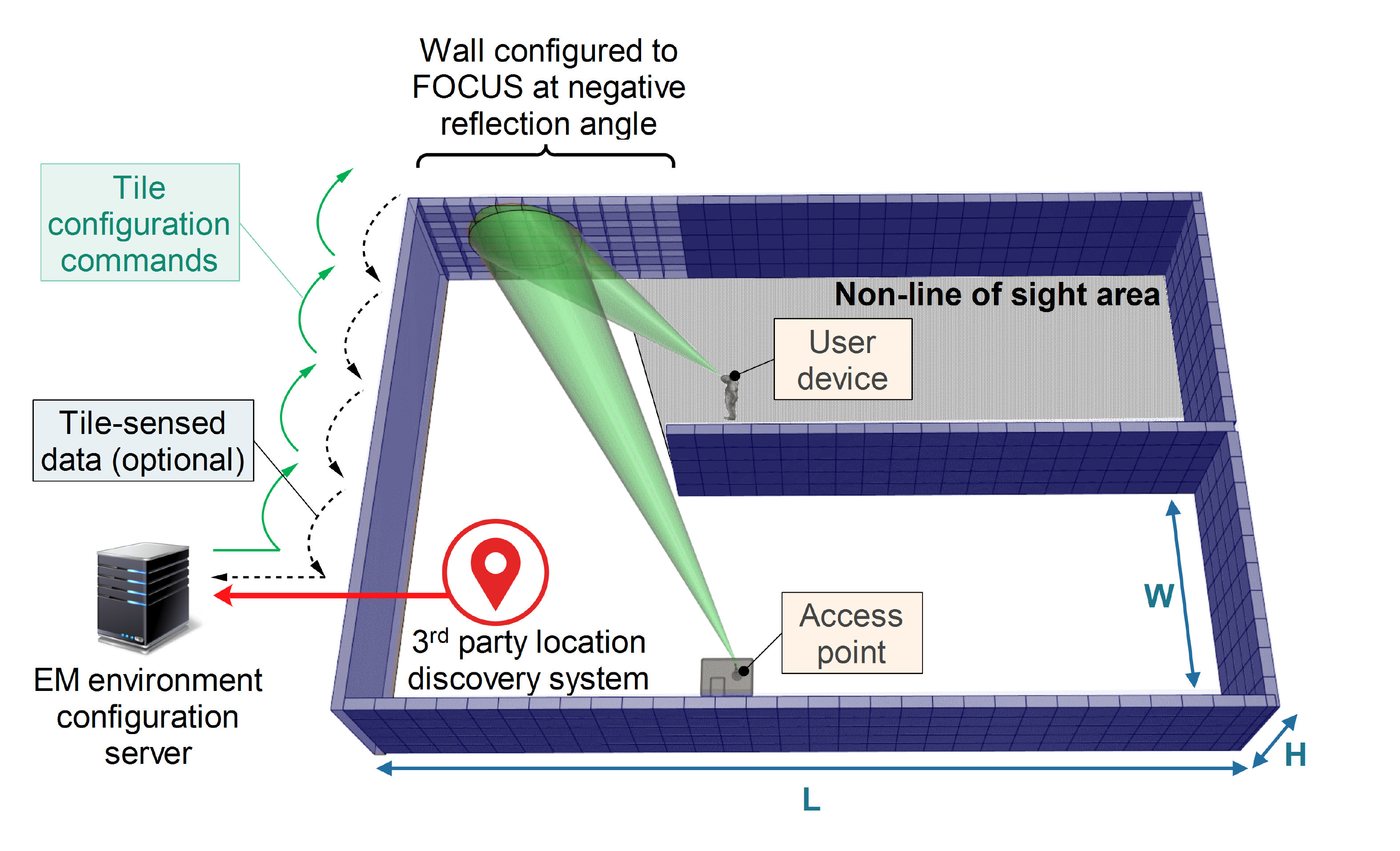}
\par\end{centering}
\caption{\label{fig:WirCommExampleNegAngle}Illustration of a customized wireless
indoor environment. $\texttt{STEER}$ functions are applied to several
tiles, to achieve a $\texttt{FOCUS}$ behavior of the corresponding
wall as a whole. }
\end{figure}

In mm-wave setups, major factors affect the signal attenuation: i)
the increased free space path loss (e.g., $\sim90\,dB$ at $10\,m$
for $60\,GHz$, instead of $60\,dB$ for $2.4\,GHz$), ii) acute multi-path
fading even in LOS cases, iii) strong Doppler shift even at pedestrian
speeds, iv) optical-like propagation of EM waves, limiting connectivity
to LOS cases and exhibiting strong sensitivity to shadowing phenomena.
Attenuation due to molecular absorption may not play a significant
role in indoor cases\textendash depending on the composition of the
environment\textendash as it corresponds to $10^{-5}\,dB/m$ loss~\cite{pi2016millimeter}.

Given the mentioned mm-wave considerations, we proceed to present
mitigation measures offered by a HyperSurface-enabled environment.
We consider the setup of Fig.~\ref{fig:WirCommExampleNegAngle},
comprising a receiver (Rx)-transmitter (Tx) pair located in NLOS over
a known floorplan. The walls are coated with HyperSurface tiles. Furthermore,
we consider the existence of a location discovery service (e.g.,~\cite{localization60ghzCmaccuracy}),
which reports the location of the user device. At first, the Rx and
Tx may attempt high-power, omni-directional communication. The location
discovery service pinpoints the location of the user device and sends
it to the EM environment configuration server. (Without loss of generality,
the location of the Tx/access point can be considered known). Tiles
may sense their impinging power and report it to the server as well.
The server can use this information to increase the accuracy of the
discovered user device location. Subsequently, the following actions
take place:
\begin{itemize}
\item The tiles at the top-left part of Fig.~\ref{fig:WirCommExampleNegAngle}
are set to a symmetric ``negative focus'' setup as shown.
\item The Tx and the Rx are signaled to direct their antenna patterns to
the configured tiles using beamforming.
\end{itemize}
Using this approach, the path loss can be even fully mitigated, since
the emitted energy is focused at the communicating end-points, rather
than scattering within the environment. This can also be of benefit
to the user device's battery lifetime, given that the redundantly
emitted power is minimized. Concerning multi-path fading, the fine-grained
EM control over the wave propagation can have as an objective the
\emph{crafting} of a power delay profile that mitigates the phenomenon,
e.g., by ensuring a path with significantly more power than any other,
or one that best matches the MIMO capabilities of the devices. Additionally,
the focal point of the EM wave reflected by the tiled wall towards
the use device can be altered in real-time, to match the velocity
of the mobile user. Mobile trajectory predictions can be employed
to facilitate this course of action. This provides a potential mitigation
approach for Doppler phenomena.

The environment optimization for multiple user pairs, or sub-spaces
within the environment, may be of increased practical interest. Returning
to the setup of Fig.~\ref{fig:WirCommExampleNegAngle}, the configuration
server can, e.g., set the tiles to preemptively minimize the delay
spread within the whole NLOS area, while ensuring a minimum level
of received power within it. In the sub-space optimization case, the
best matching tile configurations can be calculated offline and be
deployed upon request. This approach is evaluated in Section~\ref{sec:Evaluation}.

Finally, it is noted that the programmable environment extends the
communication distance of devices, without requiring extra dissipation
of energy within the environment (e.g. by placing additional access
points). This can constitute a considerable advantage for mm-wave
communications, which are known to be highly absorbable by living
tissue. Moreover, assuming tiles with state-preserving switch elements,
the energy footprint of the programmable environment can be extremely
low, especially in static or mildly changing user positions.

\section{Evaluation in $60\,GHz$ and $2.4\,GHz$ setups\label{sec:Evaluation}}

We proceed to evaluate the HyperSurface potential in mitigating the
path loss and multi\textendash path fading effects, using a simulation
platform. Specifically, the indoor 3D space of Fig.~\ref{fig:WirCommExampleNegAngle}
is ported to a full-3D ray-tracing engine~\cite{ActixLtd.2010},
customized to take into account HyperSurface tile functions. The evaluation
focuses on finding tile configurations that optimally mitigate the
path loss and multi\textendash path fading for $12$ users within
the NLOS area. We study the case of $60\,GHz$, which is of increased
interest to upcoming 5G communications, as well as the $2.4\,GHz$
case due to its wide applicability, e.g., to WiFi setups~\cite{yilmaz2016millimetre}.

Concerning the simulation parameters, the space has a height of $H=3\,m$,
corridor length (distance between opposite wall faces) $L=15\,m$,
corridor width $W=4.5\,m$, a middle wall length of $12\,m$, and
$0.5\,m$ wall thickness. Two stacked walls exist in the middle. The
floor and ceiling are treated as plain, planar surfaces composed of
concrete, without HyperSurface functionality. All walls are coated
with HyperSurface tiles, which are square-sized with dimensions $1\times1\,m$.
Thus, the 3D space comprises a total of $222$ tiles.

The dynamic metasurface pattern of Fig.~\ref{fig:Mspatterns} is
considered, using mechanical, state-preserving switches (ON/OFF states
are allowed). Appropriate dimensions are assumed, for $60\,GHz$ and
$2.4\,GHz$ respectively, as explained in the context of Fig.~\ref{fig:Mspatterns}.
This pattern design has been extensively studied in literature, offering
a wide range of steering and absorbing capabilities, even with switch
elements only at the horizontal direction~\cite[p. 235]{haupt2007genetic}.
Although beyond of the present scope, it is noted that this metasurface
design also exhibits tunable EM interaction frequency, yielding a
particularly extended repertoire of supported tunability parameters.
The considered tile functions account for EM wave steering and absorption
from various DoAs. Specifically, we allow for any DoA and reflection
direction resulting from the combination of $\left\{ -30^{o},\,-15{}^{o},\,0^{o},\,15^{o},\,30^{o}\right\} $
in azimuth and $\left\{ -30^{o},\,-15{}^{o},\,0^{o},\,15^{o},\,30^{o}\right\} $
in elevation planes, using the tile center as the origin. Notice that
the considered angles have been shown to be commonly attainable by
metasurfaces~\cite{Yazdi.2017}. However, carefully designed, static
metasurfaces have achieved nearly full angle coverage, i.e., almost
$\left(-90^{o},\,90^{o}\right)$ in azimuth and elevation, which is
indicative of their potential~\cite{Albooyeh.2014}. The reflection
coefficient is set to $100\%$ for each steering function~\cite[p. 235]{haupt2007genetic}.
Additionally, we consider an EM absorbing tile function which reduces
the power of impinging waves (given DoA) by $35\,dB$~\cite[p. 235]{haupt2007genetic},
scattering the remaining wave power towards the Snell law-derived
reflection direction. Thus, a tile supports $26$ different function
configurations in total.
\begin{figure*}[t]
\begin{centering}
\includegraphics[bb=50bp 240bp 550bp 650bp,clip,width=0.49\textwidth]{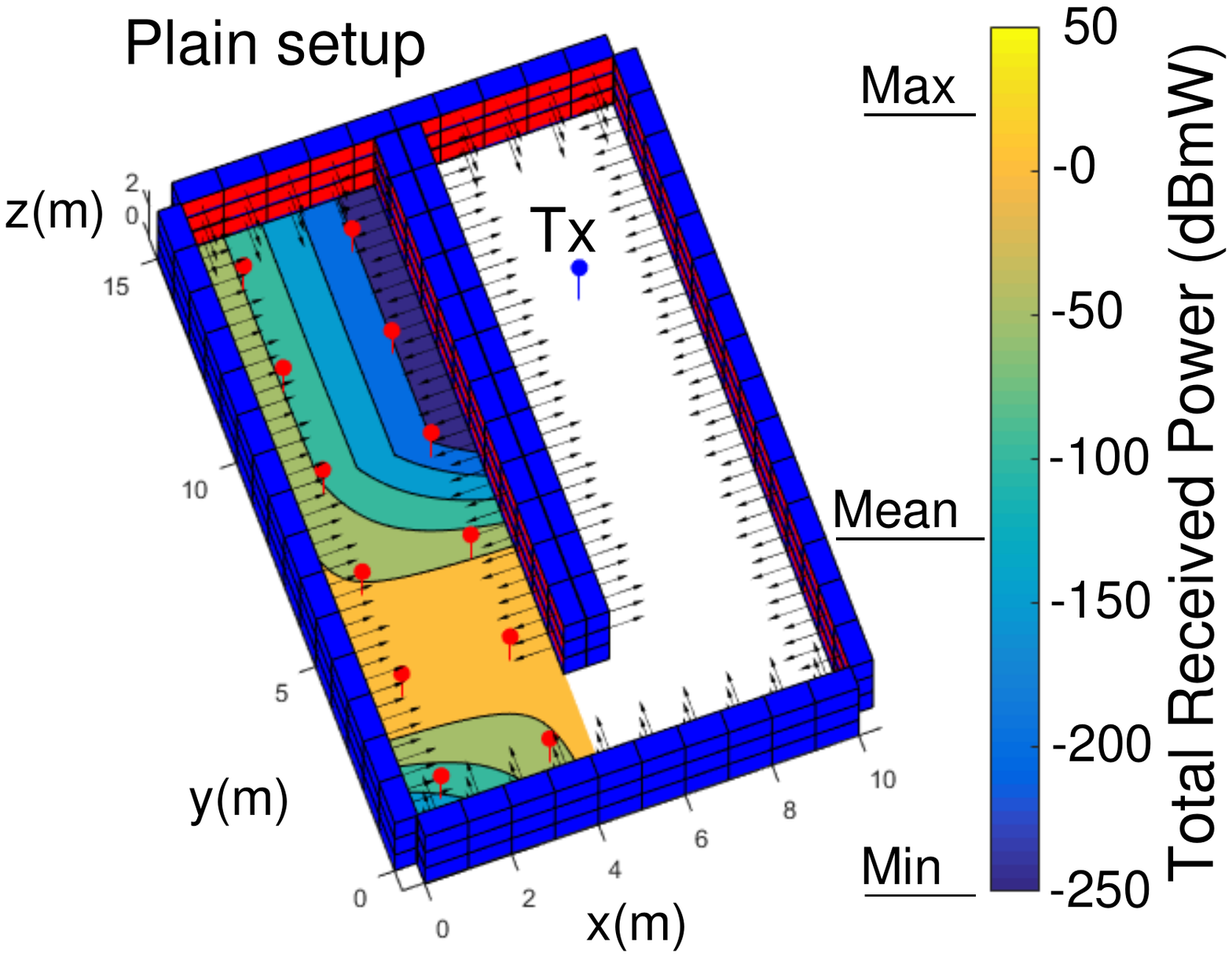}\includegraphics[bb=60bp 200bp 590bp 650bp,clip,width=0.52\textwidth]{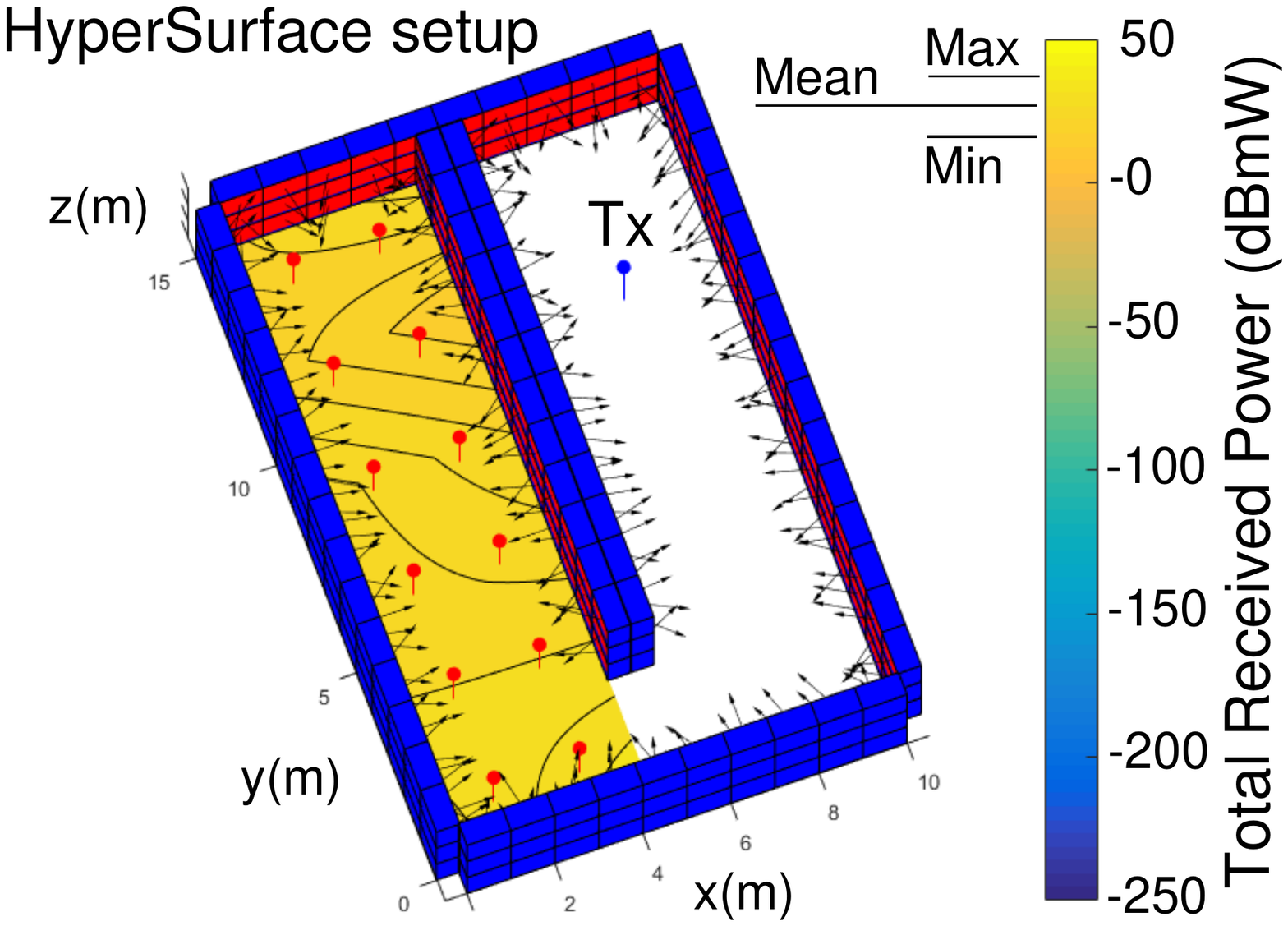}
\par\end{centering}
\caption{\label{fig:maxminP60}Wireless environment optimization case study
(A) for $60\,GHz$ and comparison to the plain case (non-HyperSurface).
The objective is to maximize the minimum total received power over
the NLOS area receivers (red dots). }
\end{figure*}

Existing ray-tracing engines employ common laws of optics to simulate
the propagation of waves. As such, current ray-tracers do not readily
allow for custom wave steering functions. (Absorbing functions, on
the other hand, are readily supported). Thus, to implement steering
functions we work as follows. First, the following observation is
made:
\begin{rem}
Assume a tile and a set of a required wave DoA and a reflection direction
upon it, not abiding by Snell's law. There exists a rotation of the
tile in 3D space that makes the wave DoA and reflection direction
comply with Snell's law.
\end{rem}
Based on this Remark, the custom steering functions are implemented
by tuning the tile's spatial derivative as follows. Since a tile is
a flat, square surface in a 3D space, its spatial derivative is normally
an arrow perpendicular to the tile surface. In order to allow for
custom EM wave steering within the ray-tracing engine, we allow for
virtually rotating the spatial derivative (but not the tile itself)
by proper azimuth and elevation angles. The modified spatial derivative
is then used in all ray-tracing calculations.

The external service is considered to know the tile specifications,
i.e., the tile configuration that corresponds to each virtual angle
combination. The service has obtained the direction of the impinging
wave at each tile via the distributed sensing elements. Subsequently,
it deploys the corresponding STEER or ABSORB commands at each tile,
by applying the corresponding tile configuration.

An EM transmitter (Tx) is placed at position $\left\{ 7,\,12,\,2\right\} \,m$
(with respect to the origin placed on the floor level, at the upper-left
corner of Fig.~\ref{fig:WirCommExampleNegAngle}). It is equipped
with a half-dipole antenna and transmits at a carrier frequency of
$60\,GHz$ or $2.4\,GHz$ (two studies) and $25\,MHz$ bandwidth.
The transmission power is set to $100\,dBmW$, a high number chosen
to ensure that no propagation paths are disregarded by the ray-tracer
due to its internal, minimum-allowed path loss threshold. The NLOS
area is defined as $x\in\left[0,\,4\right]\,m$, $y\in\left[0,\,15\right]\,m$
and a constant height of $z=1.5\,m$. Within the NLOS area, a set
$12$ receivers\textendash with antennas identical to the transmitter\textendash are
placed at a regular $2\times6$ uniform grid deployment, with $2.5\,m$
spacing. The receiver grid is centered in the NLOS area. Intermediate
signal reception values, used only for illustration purposes in the
ensuing Figures, are produced by means of interpolation.

The evaluation scenario considers two case studies, corresponding
to the path loss and multi-path fading mitigation objectives. In each
case, the state of each of the $222$ tiles is treated as an input
variable of an appropriate objective function which must be optimized.
Given the vastness and discontinuity of the solution space (i.e.,~$222^{26}$
possible tile configurations, positioned at different walls) and the
discrete nature of the input variables, a Genetic Algorithm (GA) is
chosen as the optimization heuristic~\cite{haupt2007genetic}, using
the MATLAB Optimization Toolbox implementation~\cite{MatlabGA}.
GAs are heuristics that are inspired by evolutionary biology principles.
They treat the variables of an optimization problem as \emph{genomes}
which compete with each other in terms of best fitness to an optimization
objective. Good solutions are combined iteratively by exchanging \emph{genes},
i.e., variable sub-parts, producing new generations of solutions.
In the problem at hand, a genome represents a complete tile configuration,
i.e, an array containing the state of the $222$ tiles. A gene represents
the state of each tile, i.e., the specific array elements. Two optimization
cases are studied, denoted as (A) and (B), both for $60\,GHz$ and
$2.4\,GHz$. These are defined as follows:
\begin{itemize}
\item \textbf{\uline{Case study (A)}}. This case expresses the path loss
mitigation goal, and is defined as the following optimization objective:
\emph{Define the optimal tile configurations that maximize the minimum
received power over the 12 receivers in the NLOS area}.
\item \textbf{\uline{Case study (B)}}. The case expresses the multi-path
fading mitigation goal and is defined as the following optimization
objective: \emph{Define the optimal tile configurations that minimize
the maximum delay spread over the 12 receivers in the NLOS area, with
the constraint of ensuring a minimum total received power (custom
threshold}).
\end{itemize}
For Case (B), the thresholds are set to \emph{$1\,dBmW$} for $60\,GHz$,
and \emph{$30\,dBmW$} for $2.4\,GHz$, based on the floor-plan dimensions
and the path loss levels discussed in Section~\ref{sec:App}.
\begin{table}
\caption{\textsc{\label{tab:T60}Comparison of total received power (case A)
and power delay profile (case B) with and without HyperSurface (HSF)
Tiles at $60\,GHz$.}}
\centering{}%
\begin{tabular}{|c|c|c|c|c|}
\cline{2-5}
\multicolumn{1}{c|}{} & \multicolumn{2}{c|}{Case A ($dBmW$)} & \multicolumn{2}{c|}{Case B ($nsec$)}\tabularnewline
\cline{2-5}
\multicolumn{1}{c|}{} & HSF setup  & Plain setup  & HSF setup  & Plain setup\tabularnewline
\hline
Max  & \textbf{$34.98$}  & $22.63$  & $0.69$  & $3.6$\tabularnewline
\hline
Mean  & $25.38$  & $-75$  & $0.0068$  & $0.48$\tabularnewline
\hline
Min  & $16.13$  & $-250$  & $0.0045$  & $0.007$\tabularnewline
\hline
\end{tabular}
\end{table}

\begin{figure}[t]
\begin{centering}
\includegraphics[bb=40bp 170bp 1000bp 650bp,clip,width=1\columnwidth]{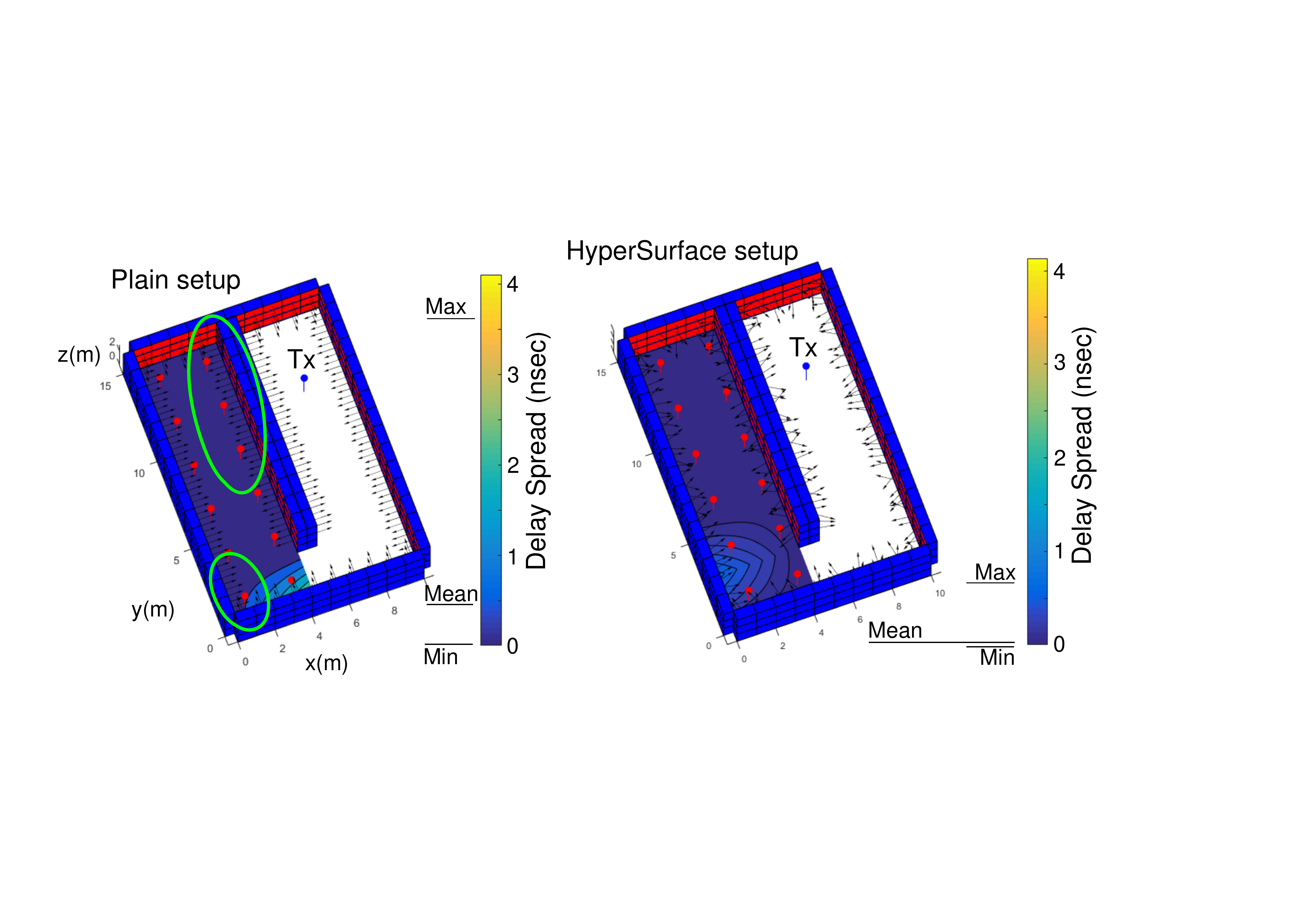}
\par\end{centering}
\caption{\label{fig:minmaxDS60}Wireless environment optimization case study
(B) for $60\,GHz$. The objective is to minimize the maximum delay
spread over the NLOS area, while ensuring a minimum of $1\,dBmW$
total received power per receiver. The circled parts of the plain
setup correspond to disconnected areas. (cf. Fig.~\ref{fig:maxminP60}-left).}
\end{figure}

The results for the $60\,GHz$ case are shown in Fig.~\ref{fig:maxminP60},
\ref{fig:minmaxDS60} and are summarized in Table.~\ref{tab:T60}.
Figure~\ref{fig:maxminP60} presents case (A) for the plain (left)
and HyperSurface-enabled (right) environments. In the plain setup,
the tile spatial derivatives (black arrows) are naturally perpendicular
to the tile surfaces. The average received power over the $12$ NLOS
area receivers is $-75\,dBmW$, while the minimum power is $-250\,dBmW$,
which is the lowest level allowed by the ray-tracing engine. Thus,
the bottom-left and the three top-right receivers of the NLOS area
are essentially disconnected in the plain setup. The maximum total
received power is $22.63\,dBmW$.

The right inset of Fig.~\ref{fig:maxminP60} shows the corresponding
results with the HyperSurface functionality enabled. Notably, the
minimum power level over the NLOS area is $16.13\,dB$, which constitutes
a raise by at least $266.13\,dBmW$ with regard to the plain case.
Moreover, the received power becomes essentially uniform over the
NLOS area, ranging between $16.13$ and $34.98\,dBmW$, with an average
of $25.38\,dBmW$. The tile spatial derivatives exhibit a degree of
directivity towards the previously disconnected area parts (e.g.,
cf. left-most wall). Moreover, the top-and bottom tiles across the
height of the walls tend to focus towards the NLOS area height. The
non-uniformity of the derivatives is in accordance with the nature
of the Genetic Algorithm, which is a very exploratory but not gradient-ascending
optimizer~\cite{Luke.2009}. This means that there exists potential
for an even better optimization result near the Genetic Algorithm-derived
solution.

The case (B) results for $60\,GHz$ are shown in Fig.~\ref{fig:minmaxDS60}.
The objective is to minimize the maximum delay spread over the $12$
NLOS receivers, under the constraint for at least $1\,dBmW$ total
received power per receiver. For the plain setup, shown in the left
inset, we note a maximum delay spread of approximately $3.6\,nsec$.
The $1\,dBmW$ minimum power constraint is of course not satisfied,
as previously shown in Fig.~\ref{fig:maxminP60}-left. The circled
areas correspond to the under-powered/disconnected NLOS area parts.
The minimum and average delay spread over the \emph{connected} areas
only are $7\,psec$ and $0.48\,nsec$ respectively. The HyperSurface-enabled
setup (right inset), achieves $5.21$ times lower maximum delay spread
($0.69\,nsec$) than the plain setup, a minimum of $4.5\,psec$ delay
spread ($1.5$ times lower), and an average of $6.8\,psec$ ($70$
times lower). This significant performance improvement is accompanied
by considerable total power levels, in the range of $\left[7.07,\,16.93\right]\,dBmW$
(average:~$10.64\,dBmW$), fulfilling the optimization constraint
of $1\,dBmW$.
\begin{figure*}[t]
\begin{centering}
\includegraphics[bb=0bp 130bp 2400bp 720bp,clip,width=1\textwidth]{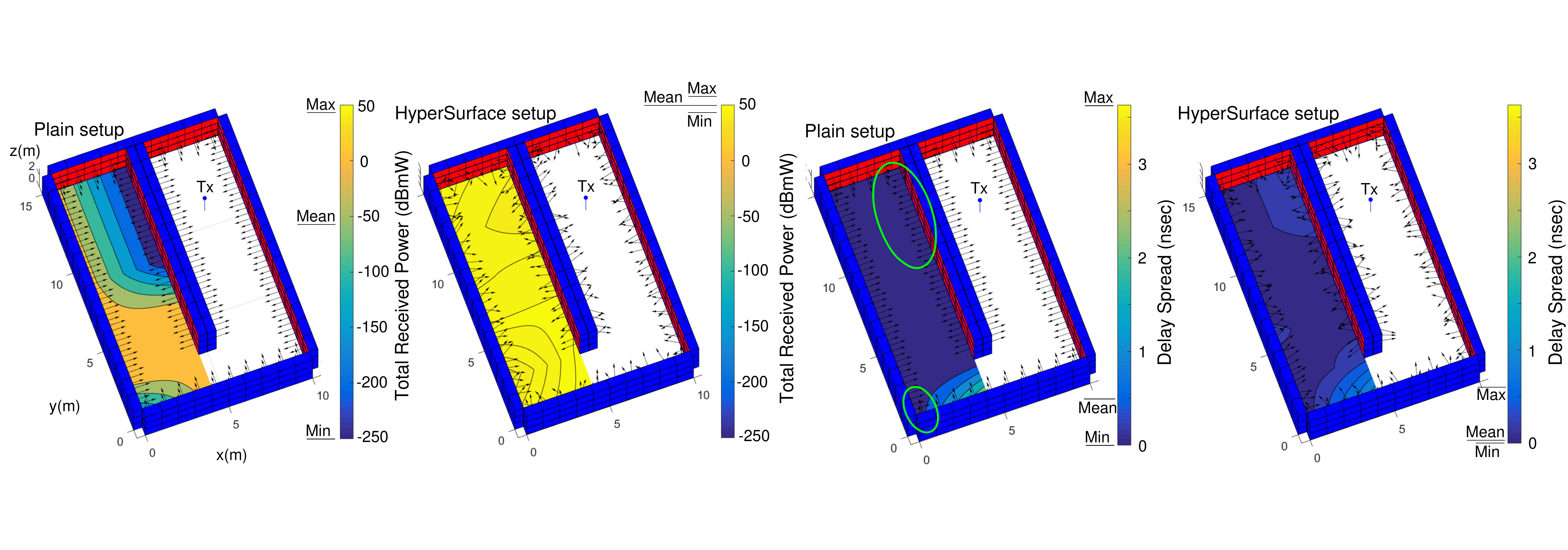}
\par\end{centering}
\caption{\label{fig:all2p4}Wireless environment optimization case studies
(A: left-two insets) and (B: right-two insets) for the $2.4\,GHz$
case. }
\end{figure*}

\begin{table}
\caption{\textsc{\label{tab:T2p4}Comparison of total received power (case
A) and power delay profile (case B) with and without HyperSurface
(HSF) Tiles at $2.4\,GHz$.}}
\centering{}%
\begin{tabular}{|c|c|c|c|c|}
\cline{2-5}
\multicolumn{1}{c|}{} & \multicolumn{2}{c|}{Case A ($dBmW$)} & \multicolumn{2}{c|}{Case B ($nsec$)}\tabularnewline
\cline{2-5}
\multicolumn{1}{c|}{} & HSF setup  & Plain setup  & HSF setup  & Plain setup\tabularnewline
\hline
Max  & \textbf{$59.81$}  & $47$  & $0.68$  & $3.65$\tabularnewline
\hline
Mean  & $51.37$  & $-58$  & $0.067$  & $0.47$\tabularnewline
\hline
Min  & $45.13$  & $-250$  & $0.0029$  & $0.0014$\tabularnewline
\hline
\end{tabular}
\end{table}

The results for the $2.4\,GHz$ case are similar to the $60\,GHz$
in terms of improvement, and are collectively given in Fig.~\ref{fig:all2p4}
and Table~\ref{tab:T2p4}. The objective in the two leftmost panels
is to maximize the minimum total received power over the $12$ receivers
in the NLOS area. The plain setup achieves $-250$, $-58$ and $47\,dBmW$
minimum, average and maximum total received power, respectively. The
HyperSurface setup yields considerably improved results, with $45.13$,
$51.37$ and $59.81\,dBmW$ minimum, average and maximum total received
power, respectively. Thus, there is a gain of $295.13\,dBmW$ in minimum
received power.

The delay spread improvement is also significant, as shown in the
two rightmost panels. The plain setup yields $1.4\,psec$, $0.47\,nsec$
and $3.65\,nsec$ minimum, average and maximum delay spread values,
with $4$ disconnected receivers (circled parts, cf. first inset of
Fig.~\ref{fig:all2p4}). The corresponding HyperSurface-enabled setup
achieves $2.9\,psec$, $67\,psec$ and $0.68\,nsec$ min/average/max
respectively. Moreover, it ensures a minimum total received power
of $34.12\,dBmW$, successfully meeting the $30\,dBmW$ optimization
constraint.

\subsection{Discussion\label{sec:future}}

The results of Section~\ref{sec:Evaluation} demonstrated the path
loss and multi-path fading mitigation potential of the proposed softwarization
of wireless indoor environments. Even at the highly-challenging $60\,GHz$
communications, a HyperSurface tile-coated indoor setup exhibited
significant improvements in received power levels and delay spread.
Such traits can benefit the communication distance of devices and
their energy consumption, without dissipating more energy in the\textendash already
EM-strained\textendash environments via retransmitters. This promising
performance can encourage further exploration of the HyperSurface
concept in additional usage domains:

Multiple applications can be studied in both indoor and outdoor environments,
and in the context of multiple systems, such as 5G, IoT and D2D, where
ultra-low latency, high bandwidth, and support for massive numbers
of devices is important~\cite{Aijaz.2017}. Moreover, HyperSurfaces
may act as an enabler for upcoming $THz$ communications. Operation
in this band promises exceptional data rates and hardware size minimization
at the nano-level, which can enable a wide range of groundbreaking
applications~\cite{akyildiz2015internet}. Nonetheless, the $THz$
band is susceptible to acute signal attenuation owed to molecular
absorption. HyperSurfaces with graphene-based meta-atom designs could
act as a smart environment for $THz$ communications~\cite{Fan.2013},
mitigating the attenuation effects and extending the communication
range.

\section{Conclusion\label{sec:Conclusion}}

The present study proposed an indoor wireless communication paradigm
where the electromagnetic propagation environment becomes aware of
the ongoing communications within it. The key idea is to coat objects
such as walls, doors and furniture with HyperSurface tiles, a forthcoming
type of material with programmable electromagnetic behavior. HyperSurfaces
can exert fine-grained control over impinging electromagnetic waves,
steering them toward completely custom directions, polarizing them
or fully absorbing them. HyperSurfaces have inter-networking capabilities,
allowing for the first time the participation of electromagnetic properties
of materials into control loops. A central server maintains a view
of the communicating devices within an indoor space, and subsequently
sets the tile electromagnetic configuration in accordance with any
optimization objective. The HyperSurface tile concept has been evaluated
in $2.4\,\text{and}\,60\,GHz$ setups, which demonstrated its high
potential for path loss and multi-path fading mitigation, from microwave
to mm-wave setups.

\section*{Acknowledgment}

This work was funded by the European Union via the Horizon
2020: Future Emerging Topics call (FETOPEN), grant EU736876, project
VISORSURF (http://www.visorsurf.eu).

The concept of Programmable Wireless Environment was introduced in the viewpoint article~\cite{cacm2018} by the respective authors.

Programmable Wireless Environments are provisionally patented by the authors of~\cite{cacm2018} as: ''A New Wireless Communication Paradigm: Realizing Wireless Environments through Software-controlled Metasurfaces'', US 62/539,831, filed 10-Aug-2017.

The present work has been published in conference form in~\cite{wowmom2018}.

A journal version of the present work appears in~\cite{wcm2018}.

\bibliographystyle{IEEEtran}

\end{document}